\documentclass[11pt,twoside]{article}

\usepackage{asp2006}
\usepackage{epsf}
\usepackage{lscape}
\usepackage{graphicx}
\usepackage{natbib}

\begin{document}
\title{The Milky Way is an Exceptionally Quiet galaxy: Implications
  for spiral formation}   
\author{M. Puech$^{1,2}$, F. Hammer$^{2}$, L. Chemin$^{2}$, H.
  Flores$^{2}$, and M. Lehnert$^{2}$}   
\affil{$^{1}$ESO,
  Karl-Schwarzschild-Strasse 2, D-85748 Garching bei M\"unchen,
  Germany\\ $^{2}$GEPI, Observatoire de Paris, CNRS, University Paris
  Diderot; 5 Place Jules Janssen, 92190 Meudon, France\\}    

\begin{abstract}
We compare both the Milky Way and M31 to local external disk galaxies
within the same mass range, using their relative locations in the
planes formed by V$_{flat}$ vs. M$_K$ (the Tully-Fisher relation),
j$_{disk}$ (specific angular momentum) and the average Fe abundance of
stars in the galaxy outskirts. We find, for all relationships, that
the MW is systematically offset by 1 $\sigma$ or more, showing a
significant deficiency in stellar mass, angular momentum, disk radius
and [Fe/H] in the stars in its outskirts at a given V$_{flat}$. Our
Galaxy appears to have escaped any significant merger over the last
10-11 Gyr which may explain its peculiar properties. As with M31, most
local spirals show evidence for a history shaped mainly by relatively
recent merging.
\end{abstract}

\section{The MW is a very peculiar galaxy}

\begin{figure}[!h]
\centering
\includegraphics[width=12cm]{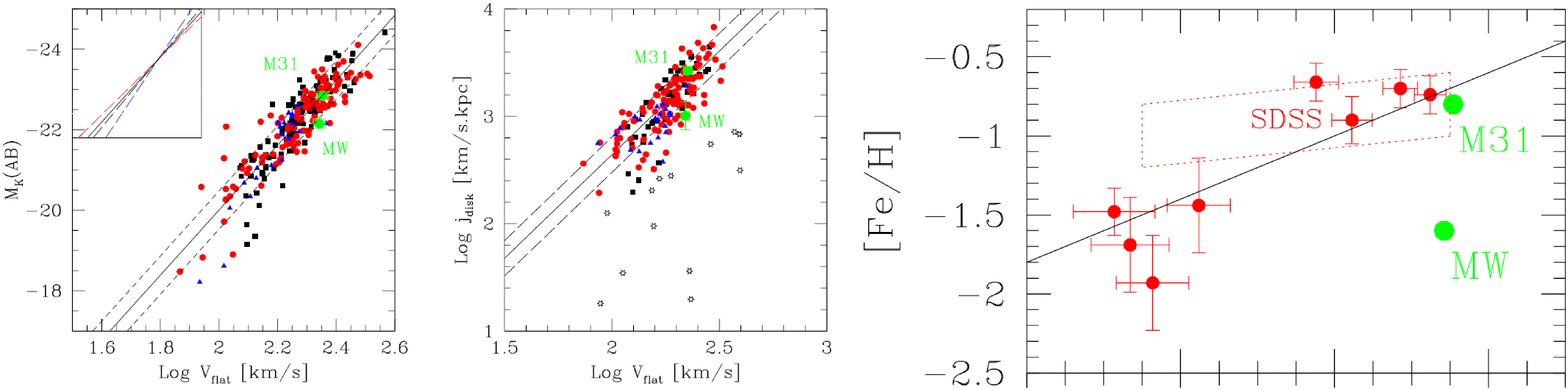}
\caption{\emph{Left and middle panels:} K-band TFR and j$_{disk}$ vs.
  V$_{flat}$ relation for three local galaxy samples (blue triangles
  from \citealt{ver01}, red dots from \citealt{pizagno07}, and black
  squares from \citealt{courteau97}). Solid lines show the fits for
  the combined three samples, and the dashed lines the residual 1
  $\sigma$ dispersion. The locations of the MW and M31 are marked by
  two large green dots. In the top-left panel insert, we show the
  individual fits of the TF relations for each of the three samples
  (see more details in \citealt{hammer07}). In the middle panel, open
  stars show results of simulations by \cite{steinmetz99},
  illustrating the difficulty in reproducing the angular momentum of
  spiral galaxies when using the standard model for disk formation.
  \emph{Right panel:} Iron abundances of stellar haloes of eight
  spiral galaxies (from \citealt{mouhcine06}) vs. log(V$_{flat}$)
  (small red points). Large green points mark the position of the
  Milky Way and M31. The full line assumes M$_{star}\sim$V$^4$
  \citep{McGaugh05} and M$_{star}\sim$Z$^2$ following the prescription
  of \cite{dekel87}. Red dashed lines identify a box including the
  color measurements of \cite{zibetti04}, after stacking 1047 edge-on
  SDSS galaxies, and assuming that their colors are dominated by red
  giant stars.}
\label{tfjfe}
\end{figure}

In Fig. \ref{tfjfe}, we show the local Tully-Fisher Relation (TFR) in
K-band, as well as the specific angular momentum of the disk
j$_{disk}$ vs. the rotational velocity V$_{flat}$. In both relations,
V$_{flat}$ is taken as a proxy for the total mass. Noteworthy, the
local TFR was carefully derived comparing three different local
samples (see top-left inset and \citealt{hammer07}). Strikingly, the
MW systematically falls at $\sim$ 1 $\sigma$ of these relations,
conversely to M31: the MW has a smaller stellar mass and specific
angular momentum, compared to local galaxies of similar total mass. On
the basis of their location in the (M$_K$, V$_{flat}$, R$_d$) volume,
the fraction of spirals like the MW is $\sim$7\%, while M31 appears to
be a much more typical spiral (see \citealt{hammer07}). The
representativity of the MW as a typical spiral galaxy is actually even
smaller, if one takes into account the iron abundance of the stellar
halo (see right panel).

\section{Implication for spiral formation}
One can explain the differences between the MW and M31 by the absence
of significant merger events over the last 10-11 Gyr for the former.
On the contrary, M31 has been much more influenced by mergers (and
possibly a major merger) during the last 8 Gyrs (see
\citealt{hammer07} for a detailed discussion). Could the bulk of local
spirals be shaped and assembled through a rich past merger history
during the last 8 Gyrs? Such a scenario is indeed suggested by direct
observations of violent dynamical processes at work in z$\sim$0.6
galaxies \citep{yang07}, and by the relatively important major merger
rate per M* galaxy during the last 11 Gyrs (0.5-0.7 up to z$\sim$1,
see \citealt{hammer07}, and 4-5 up to z$\sim$3, see
\citealt{conselice03}). In this context, rich major merger episodes in
the past may explain the Tully-Fisher relation, the angular momentum
problem and the abundances of stars in spiral outskirts. What remains
unclear is whether or not these properties could be explained by the
standard scenario for spiral formation, i.e., a monolithic collapse
followed by pure secular evolution.

\acknowledgements 
We wish to thank the organizing committees for this very pleasant meeting.


\end{document}